\documentclass[aps,prl,showpacs,floatfix,twocolumn]{revtex4-2}  
\usepackage{bm}
\usepackage{latexsym}
\usepackage{dcolumn}
\usepackage{amsfonts,amssymb,amsmath}
\usepackage{graphicx,epsfig}
\usepackage{psfrag}
\usepackage{comment}
\usepackage{url}
\usepackage{feynmf}
\usepackage{rotating}
\usepackage{hyperref}
\usepackage{enumitem}
\usepackage{xcolor}
\usepackage{multirow}
\hypersetup{
     unicode=false,          
     pdftoolbar=true,        
     pdfmenubar=true,        
     pdffitwindow=false,     
     pdfstartview={FitH},    
     pdftitle={My title},    
     pdfauthor={Author},     
     pdfsubject={Subject},   
     pdfcreator={Creator},   
     pdfproducer={Producer}, 
     pdfkeywords={keyword1} {key2} {key3}, 
     pdfnewwindow=true,      
     colorlinks=true,        
     linkcolor=red,          
     citecolor=blue,         
     filecolor=magenta,      
     urlcolor=blue,         
     linktocpage=true
}

\def\beq{\begin{equation}}
\def\eeq{\end{equation}}
\def\br{\begin{eqnarray}}
\def\er{\end{eqnarray}}
\def\benu{\begin{enumerate}}
\def\efnu{\end{enumerate}}

\begin{document}




\title{Discordances in cosmology and the violation of slow-roll inflationary dynamics}

\author{Akhil Antony} \email{akhilantony@imsc.res.in}
\affiliation{The Institute of Mathematical Sciences, HBNI, CIT Campus, Chennai 600113, India\\Homi Bhabha National Institute, Training School Complex, Anushakti Nagar, Mumbai 400085, India.}
\author{Fabio Finelli}\email{fabio.finelli@inaf.it}
\affiliation{INAF/OAS Bologna, Osservatorio di Astrofisica e Scienza dello Spazio, Area della ricerca CNR-INAF, via Gobetti 101, I-40129 Bologna, Italy\\INFN,  Sezione  di  Bologna,  via  Irnerio  46,  40126  Bologna,  Italy}
\author{Dhiraj Kumar Hazra} \email{dhiraj@imsc.res.in}
\affiliation{The Institute of Mathematical Sciences, HBNI, CIT Campus, Chennai 600113, India\\Homi Bhabha National Institute, Training School Complex, Anushakti Nagar, Mumbai 400085, India\\INAF/OAS Bologna, Osservatorio di Astrofisica e Scienza dello Spazio, Area della ricerca CNR-INAF, via Gobetti 101, I-40129 Bologna, Italy}
\author{Arman Shafieloo}\email{shafieloo@kasi.re.kr}
\affiliation{Korea Astronomy  and Space Science Institute, Daejeon 34055, Korea\\ University of Science and Technology, Daejeon 34113, Korea}
\date{\today}

\begin{abstract}
 We identify examples of single field inflationary trajectories beyond the
slow-roll regime which improve the fit to Planck 2018 data compared to baseline $\Lambda$CDM model with power law form of primordial spectrum
and at the same time alleviate existing tensions between different data sets
in the estimate of cosmological parameters such as $H_0$ and $S_8$. A damped oscillation in the first Hubble flow
function
- or equivalently a feature in the potential -
and the corresponding localized oscillations in the primordial power spectrum partially
mimic the improvement in the fit of Planck data due to $A_L$ or $\Omega_K$.
Compared to the baseline model, this model can lead {\em simultaneously} to larger value of $H_0$ and a smaller value of $S_8$,
a trend which can be enhanced when the most recent
SH0ES measurement for $H_0$ is combined with Planck and BK18 data. Large scale structure data and more precise CMB polarization measurements will further provide critical tests of this intermediate fast roll phase.

\end{abstract}

\pacs{98.80.Cq}
\maketitle


\paragraph{\textit{Introduction}:} 

Thanks to the precise measurement of the Cosmic Microwave Background (CMB) anisotropy pattern, cosmology has entered a precision era. Although $\Lambda$CDM is still theoretically incomplete in terms of the fundamental constituents of dark matter and dark energy,  
new physics beyond the six parameters of the flat $\Lambda$CDM concordance cosmology have not emerged yet 
within Planck 2018 data \cite{Planck:2018param,Planck:2018inf}. 
Among the so called “anomalies”, Planck temperature data seem to indicate an extra smoothing of the acoustic peaks at multipoles higher than 800 and a low amplitude at multipoles lower than 40 \textit{w.r.t} what $\Lambda$CDM predicts. 
These effects are the main reason why 
a phenomenological rescaling of the lensing amplitude $A_{\rm L}$ or a positive spatial 
curvature $\Omega_K$~\cite{Planck:2018param} improve the fit for Planck temperature and polarization data at nearly 3$\sigma$, although 
the stretches in these parameters are reduced when Planck CMB lensing is included in the analysis.

Two theoretical explanations for this extra smoothing, such as compensated CDM isocurvature perturbations and 
primordial features were proposed in \cite{Planck:2018inf}. The latter consisted an analytical template for superimposed oscillations linear in the wavenumber with a Gaussian envelope, similar to the form that can mimic CMB lensing obtained earlier in the blind reconstruction~\cite{HazraMRLPlanck:2014}, that
can mimic the smoothing of the acoustic peaks in a similar way to $A_{\rm L}$ for temperature anisotropies. Whereas the amplitude of compensated CDM isocurvature was drastically reduced when the Planck CMB lensing is added~\cite{Planck:2018inf}. Reconstruction and analytical templates of primordial features mimicking $A_{\rm L}$ were further studied in~\cite{Hazra:2022rdl}.

In this~\textit{letter}, we connect inflationary dynamics beyond slow-roll to the localized superimposed oscillations which provide an improved fit to Planck data by employing 
a profile in the Hubble parameter during inflation. 
Equipped by this theoretical proposal we then study how these particular types of localized 
features which could be hidden in Planck lead simultaneously to a larger $H_0$ and a lower $S_8$, and therefore potentially alleviate existing tensions of CMB data with Cepheid calibrated SNIa and galaxy clustering and weak lensing data. Other proposals to increase $H_0$ such as Early dark energy~\cite{PoulinPRL} or some model of scalar-tensor gravity~\cite{Rossi:2019lgt} (see \cite{DiValentino:2020zio,Schoneberg:2021qvd} for reviews), also increases $S_8$ that worsen the tension with galaxy weak lensing data. 
Our work here represents the first inflationary solution leading {\em simultaneously} to higher $H_0$ and lower $S_8$. We also derive the class of inflaton potential from the Hubble parameter as new theoretical models of concordance.


\paragraph{\textit{Model, methodology, data and analysis}:}

In order to reconstruct the Hubble flow parameters, we assume a baseline parametrization as follows, 
\begin{equation}
    \epsilon_H^{baseline} (N) = \epsilon_1 \exp\left[\epsilon_2(N-N_*)\right]~\label{eq:baseline}
\end{equation}
$N_*$ is chosen as the \textit{e-fold} at which the pivot scale ($0.05~Mpc^{- 1}$) crosses the Hubble radius. 
In this parametrization, we obtain the spectral tilt to be $n_s \simeq 1-2\epsilon_1-\epsilon_2$ and the tensor to scalar ratio to be $r\simeq 16\epsilon_1$. 
This correspondence 
allows uncertainties on $n_s$ and $r$ 
similar to those obtained by sampling directly on physical parameters (see \cite{Finelli:2009bs,Planck:2013jfk} for an analogous result also by taking into account the running ).
In this slow roll regime, we introduce an intermediate fast roll phase. The fast roll phase proposed here is modelled by sinusoidal oscillations that are damped by an envelope. Thus, the evolution of the Hubble slow roll parameter in the full model is given as,
\begin{equation}
    \epsilon_H(N) = \epsilon_H^{baseline} (N) \left( 1+ \frac{\alpha\cos\left[\omega(N-N_0)\right]}{1 + \beta(N-N_0)^2}\right),~\label{eq:feature}
\end{equation}
where the term in the parenthesis denotes the fractional change in the slow roll dynamics. The oscillation amplitude parameter is denoted by $\alpha$. The decaying oscillations peak at $N=N_0$ with a frequency $\omega$. The decay parameter is denoted as $\beta$. With $\beta\to0$, this evolution generates resonant features~\cite{Chen:Osc000,Flauger:Osc1,Chen:2010folded,Aich:Osc3} in the primordial power spectrum, while for $\omega\to0$ case is similar to having a step in the inflationary potential~\cite{Adams:2001Step,Hazra:2010Step,Hazra:2014WWI,Chluba:2015Step,Ballardini:2016hpi,Hazra:2017WWI,Hazra:2021WWI} which produces sharp features. In this general form, this flow function generates an envelope of $\sin+\sin\log+\sin$ oscillations where at $N\to N_0$, the $\sin\log$ part is dominant (similar to~\cite{Antony:2021a}).

For the inflationary dynamics defined by~\autoref{eq:baseline} and~\autoref{eq:feature} we numerically solve for the Hubble parameter as a function of \textit{e-folds}. We solve the Mukhanov-Sasaki equation
from the time when the modes are deep within the Hubble radius (imposing Bunch-Davies vacuum initial conditions) to the super Hubble scales. Both scalars and tensor perturbations are computed in this method. Note that in this formalism, the scalar potential and the evolution of the scalar field are not explicitly used and only the form of Hubble flow function remains sufficient for the estimation of scalar and tensor spectra. To solve for the Hubble parameter, we need the initial value of Hubble parameter $H_i$ at $N=0$. We use $H_i^2/\epsilon_1$ as the free parameter since it is proportional to the scalar spectral amplitude. Therefore for baseline analysis we have three free parameters in the perturbation sector, namely $H^2/\epsilon_1$, $\epsilon_1$ and $\epsilon_2$. In the paper containing the detailed analysis~\cite{Antony:2022b}, we show that constraints on our baseline model are nearly indistinguishable with the constraints obtained on scalar spectral amplitude, tilt and tensor-to-scalar ratio in the power law form of primordial spectrum. When~\autoref{eq:feature} is used for inflationary dynamics, we allow the fast roll parameters to vary alongside the baseline parameters. There exist internal degeneracies within these parameter. We do not expect Planck CMB constraints beyond the cosmological scales corresponding to $\ell\sim2500$. We find that with higher $\omega$, higher $\alpha$ reproduces the same spectrum till the Planck probed scales. Also, for high values of $\beta$, the power spectrum exhibits oscillations with a frequency independent of $\beta$ and remains degenerate with the amplitude parameter. Due to this degeneracy, we find that this decay parameter cannot be constrained beyond a lower bound. We therefore fix this parameter to $\log_{10}\beta=2.5$. Note that, fixing this parameter does not change the significance of the results noticeably. Therefore, in the feature part, we vary $\alpha,\omega$ and $N_0$. 


We modified the code BINGO~\cite{BINGO:2013} to incorporate the Hubble flow function instead of inflationary potential and use it as an add-on to CAMB~\cite{CAMB}. In order to capture the entire primordial feature in the angular power spectrum, we compute the angular power spectrum at every multipole and avoid any interpolation. CosmoMC~\cite{CosmoMC} used for parameter significance estimation.

In terms of CMB data we use the Planck 2018 ~\cite{Planck:2019Like} and BICEP/Kec k 2018 \cite{BICEP:2021xfz} data. We use four Planck dataset combinations, i.e. TT+lowl+lowE+BK18 (denoted as P18TT+BK18),  TTTEEE+lowl+lowE+BK18 (denoted as P18TP+BK18),  TTTEEE+lowl+lowE+lensing+BK18 (denoted as P18TPL+BK18) and TEEE+lowl+lowE+BK18 (denoted as P18TEEE+BK18). For high-$\ell$ TT, TEEE, TTTEEE we restrict ourselves to Plik binned datasets, since the features that help in resolving the concordance problem do not have very high frequency oscillations~\cite{Hazra:2022rdl} (in certain cases we have also tested our models with unbinned Plik datasets and found nearly identical results).
We allow all the foreground and calibration parameters corresponding to the datasets to vary as fast parameters. BK18 likelihood helps in constraining $\epsilon_1$. 
We also use the recently released $H_0$ measurement from SH0ES~\cite{Riess:2021jrx} (denoted as S21) with the above four data combinations separately.

\paragraph{\textit{Results}:}

\begin{table*}[]
\begin{tabular}{|c|clc|c|c|c|c|c|c|}
\hline
\multicolumn{1}{|c|}{\multirow{2}{*}{Data}} & \multicolumn{3}{c|}{$\Delta\chi^2$}                  & \multirow{2}{*}{C.L. } & \multicolumn{1}{c|}{\multirow{2}{*}{\begin{tabular}[c]{@{}c@{}}$1-2\epsilon_1-\epsilon_2 $\\ $(\simeq n_s)$\end{tabular}}} & \multicolumn{1}{c|}{\multirow{2}{*}{\begin{tabular}[c]{@{}c@{}}$16\epsilon_1 $\\ $(\simeq r)$\end{tabular}}} & \multicolumn{1}{c|}{\multirow{2}{*}{$H_0$}} & \multicolumn{1}{c|}{\multirow{2}{*}{$S_8$}} & \multicolumn{1}{c|}{\multirow{2}{*}{$\Omega_m$}} \\ \cline{2-4}
\multicolumn{1}{|c|}{}                      & \multicolumn{1}{c|}{Total} & \multicolumn{1}{c|}{CMB}  &  SH0ES &                       & \multicolumn{1}{c|}{}                    & \multicolumn{1}{c|}{}                   & \multicolumn{1}{c|}{}                      & \multicolumn{1}{c|}{}                    & \multicolumn{1}{c|}{}                                           \\ \hline
\multicolumn{1}{|c|}{\multirow{2}{*}{P18TT+BK18}}  & \multicolumn{1}{c|}{\multirow{2}{*}{-8.3}}   & \multicolumn{1}{c|}{\multirow{2}{*}{-8.3}}  &\multirow{2}{*}{-}   & \multirow{2}{*}{82.7}                 & 0.963$\pm$0.005  & $<$0.036  & 66.86$\pm$0.86   &  0.840$\pm$0.022  & 0.321$\pm$0.012       \\ \cline{6-10}
\multicolumn{1}{|c|}{}  & \multicolumn{1}{c|}{}   & \multicolumn{1}{c|}{}  &   &                 &\colorbox{lightgray}{0.971$\pm$0.007}  &\colorbox{lightgray}{$<$0.040}  & \colorbox{lightgray}{68.06$\pm$1.14}   & \colorbox{lightgray}{0.814$\pm$0.027}  & \colorbox{lightgray}{0.306$\pm$0.015}       \\ \hline

\multicolumn{1}{|c|}{\multirow{2}{*}{P18TEEE+BK18}}  & \multicolumn{1}{c|}{\multirow{2}{*}{-2.7}}   & \multicolumn{1}{c|}{\multirow{2}{*}{-2.7}}  &\multirow{2}{*}{-}   & \multirow{2}{*}{$<68$}                 & 0.969$\pm$0.009  & $<$0.041  & 67.91$\pm$0.77   &  0.814$\pm$0.020  & 0.308$\pm$0.010       \\ \cline{6-10}
\multicolumn{1}{|c|}{}  & \multicolumn{1}{c|}{}   & \multicolumn{1}{c|}{}  &   &                 &\colorbox{lightgray}{0.968$\pm$0.009}  &\colorbox{lightgray}{$<$0.041}  & \colorbox{lightgray}{67.63$\pm$0.86}   & \colorbox{lightgray}{0.819$\pm$0.022}  & \colorbox{lightgray}{0.311$\pm$0.012}       \\ \hline

\multicolumn{1}{|c|}{\multirow{2}{*}{P18TP+BK18}}          & \multicolumn{1}{c|}{\multirow{2}{*}{-10.7}}  & \multicolumn{1}{c|}{\multirow{2}{*}{-10.7}} &  \multirow{2}{*}{-}    &  \multirow{2}{*}{72.5}                  & {0.965}$\pm$0.004  & $<$0.036  & {67.26}$\pm$0.59   & {0.835}$\pm$0.015  &{0.317}$\pm$0.008      \\ \cline{6-10}
\multicolumn{1}{|c|}{}          & \multicolumn{1}{c|}{}  & \multicolumn{1}{c|}{} & \multicolumn{1}{c|}{}  &      &\colorbox{lightgray} {0.969$\pm$0.005}  & \colorbox{lightgray}{$<$0.037}  & \colorbox{lightgray}{67.71$\pm$0.66}   & \colorbox{lightgray}{0.826$\pm$0.017}  &\colorbox{lightgray}{0.311$\pm$0.009}      \\ \hline

\multicolumn{1}{|c|}{\multirow{2}{*}{P18TPL+BK18}} & \multicolumn{1}{c|}{\multirow{2}{*}{-8.4}}   & \multicolumn{1}{c|}{\multirow{2}{*}{-8.4}}  & \multirow{2}{*}{-}   & \multirow{2}{*}{70}               & {0.965}$\pm$0.004  & $<$0.035  & {67.35}$\pm$0.53   &{0.832}$\pm$0.012  & {0.315}$\pm$0.007       \\ \cline{6-10}
\multicolumn{1}{|c|}{} & \multicolumn{1}{c|}{}   & \multicolumn{1}{c|}{}  &   &                    & \colorbox{lightgray}{0.968$\pm$0.004}  & \colorbox{lightgray}{$<$0.037}  & \colorbox{lightgray}{67.63$\pm$0.57}   &\colorbox{lightgray}{0.829$\pm$0.013}  & \colorbox{lightgray}{0.312$\pm$0.008}       \\ \hline

\hline

\multicolumn{1}{|c|}{\multirow{2}{*}{P18TT+BK18+S21}}   & \multicolumn{1}{c|}{\multirow{2}{*}{-19.5}}  & \multicolumn{1}{c|}{\multirow{2}{*}{-10.9}} & \multirow{2}{*}{-8.6} & \multirow{2}{*}{$>99.9$}   & {0.976}$\pm$0.005  & $<$0.040   & {69.41}$\pm$0.68   & {0.781}$\pm$0.017  & {0.287}$\pm$0.008       \\ \cline{6-10}
\multicolumn{1}{|c|}{}  & \multicolumn{1}{c|}{}   & \multicolumn{1}{c|}{}  &   &  & \colorbox{lightgray}{0.986$\pm$0.007}  & \colorbox{lightgray}{$<$0.047}   & \colorbox{lightgray}{70.85$\pm$0.78}   & \colorbox{lightgray}{0.754$\pm$0.018}  & \colorbox{lightgray}{0.273$\pm$0.008}       \\\hline

\multicolumn{1}{|c|}{\multirow{2}{*}{P18TEEE+BK18+S21}}  & \multicolumn{1}{c|}{\multirow{2}{*}{-1.2}}   & \multicolumn{1}{c|}{\multirow{2}{*}{-1.0}}  &\multirow{2}{*}{-0.2}   & \multirow{2}{*}{$<68$}                 & 0.981$\pm$0.008  & $<$0.046  & 69.76$\pm$0.63   &  0.772$\pm$0.016  & 0.284$\pm$0.007       \\ \cline{6-10}
\multicolumn{1}{|c|}{}  & \multicolumn{1}{c|}{}   & \multicolumn{1}{c|}{}  &   &                 &\colorbox{lightgray}{0.979$\pm$0.009}  &\colorbox{lightgray}{$<$0.040}  & \colorbox{lightgray}{69.77$\pm$0.67}   & \colorbox{lightgray}{0.771$\pm$0.017}  & \colorbox{lightgray}{0.284$\pm$0.008}       \\ \hline

\multicolumn{1}{|c|}{\multirow{2}{*}{P18TP+BK18+S21}}& \multicolumn{1}{c|}{\multirow{2}{*}{-19.3}}  & \multicolumn{1}{c|}{\multirow{2}{*}{-9.7}}  &\multirow{2}{*}{-9.6} & \multirow{2}{*}{98.6}  & {0.973}$\pm$0.004  & $<$0.039  & {68.71}$\pm$0.53  & {0.802}$\pm$0.014  & {0.297}$\pm$0.007       \\\cline{6-10}
\multicolumn{1}{|c|}{}  & \multicolumn{1}{c|}{}   & \multicolumn{1}{c|}{}  &   &    &\colorbox{lightgray}{0.978$\pm$0.004}  & \colorbox{lightgray}{$<$0.041}  & \colorbox{lightgray}{69.27$\pm$0.58}   & \colorbox{lightgray}{0.791$\pm$0.014}  & \colorbox{lightgray}{0.291$\pm$0.007}       \\ \hline

\multicolumn{1}{|c|}{\multirow{2}{*}{P18TPL+BK18+S21}} & \multicolumn{1}{c|}{\multirow{2}{*}{-11.5}}  & \multicolumn{1}{c|}{\multirow{2}{*}{-10.4}} & \multirow{2}{*}{-1.1} & \multirow{2}{*}{92.1}    & {0.972}$\pm$0.004  & $<$0.038  & {68.56}$\pm$0.48   & {0.808}$\pm$0.011  & {0.299}$\pm$0.006       \\ \cline{6-10}
\multicolumn{1}{|c|}{}  & \multicolumn{1}{c|}{}   & \multicolumn{1}{c|}{}  &   &                 & \colorbox{lightgray}{0.975$\pm$0.004} & \colorbox{lightgray}{$<$0.041}  & \colorbox{lightgray}{68.90$\pm$0.51}   & \colorbox{lightgray}{0.804$\pm$0.011}  & \colorbox{lightgray}{0.296$\pm$0.006}\\ \hline

\end{tabular}
\caption{Results from 8 major data combinations from our analysis. $\Delta\chi^2$ values and their breakdown are quoted \textit{w.r.t.} the baseline best fits. We also provide the marginalized confidence limits (C.L.) of the $\alpha$ parameter (in percent) deviating from the baseline $\alpha=0$ value. The results for baseline are plotted in the top row of each data combination while the results corresponding to~\autoref{eq:feature} are plotted at the bottom in grey cells. Bounds on the spectral tilt and the tensor-to-scalar ratios are provided using the parameters in the Hubble flow function assuming first order approximations. In all cases the two sided bounds represent 68\% C.L. and upper bounds (as in the case of $16\epsilon_1(\simeq r)$) represent 95\% C.L. Without the addition of $H_0$ priors, the mean value of $H_0$ increases and $S_8$ and $\Omega_m$ decreases with fast roll. We find the change is maximum in the temperature data. When polarization and lensing data are added, the extent of these shifts decreases.}\label{tab:chisq}
\end{table*}

\begin{figure*}[!htb]
\centering
\includegraphics[width=0.49\textwidth]{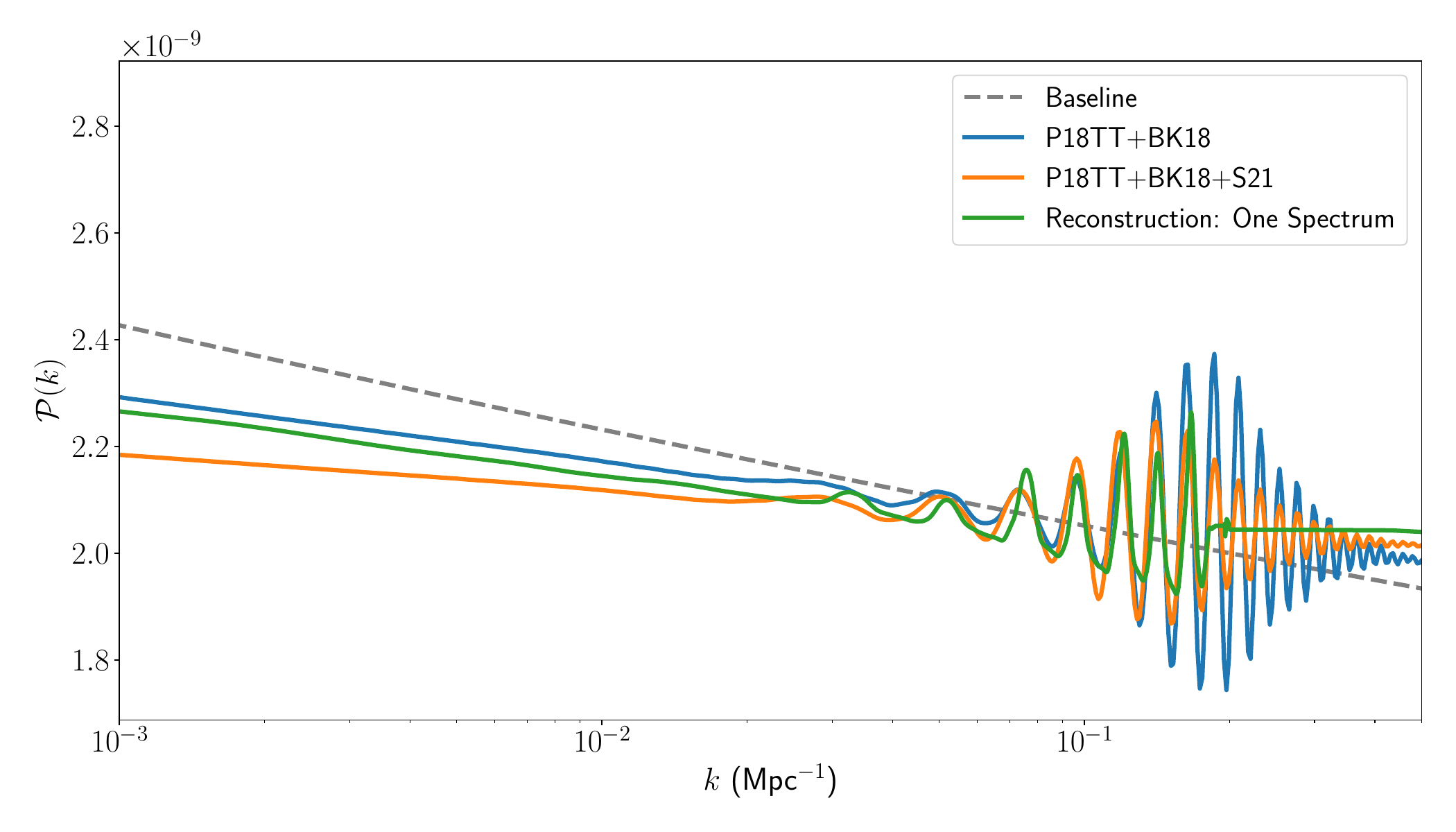}
\includegraphics[width=\columnwidth]{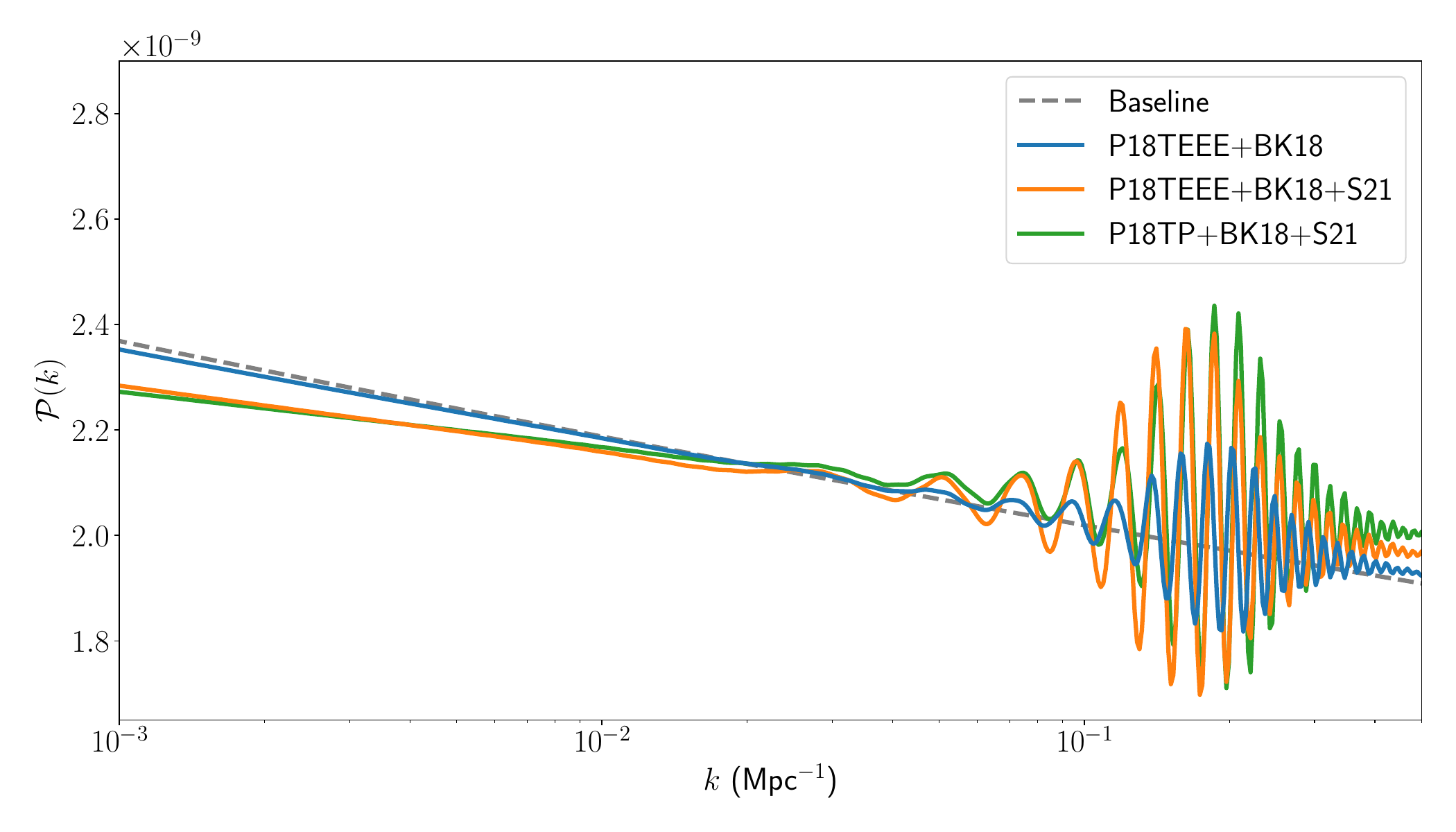}
\includegraphics[width=0.49\textwidth]{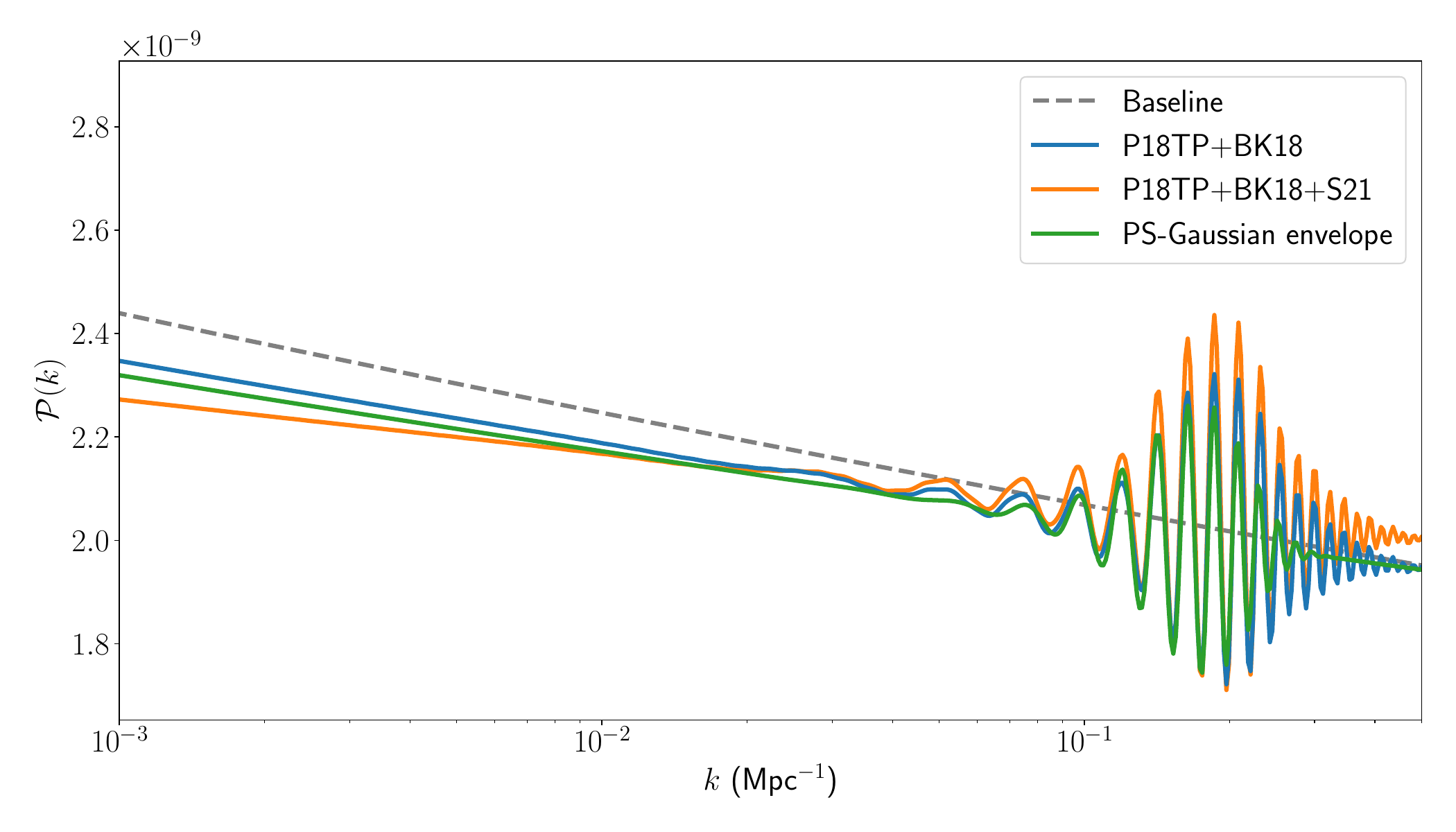}
\includegraphics[width=\columnwidth]{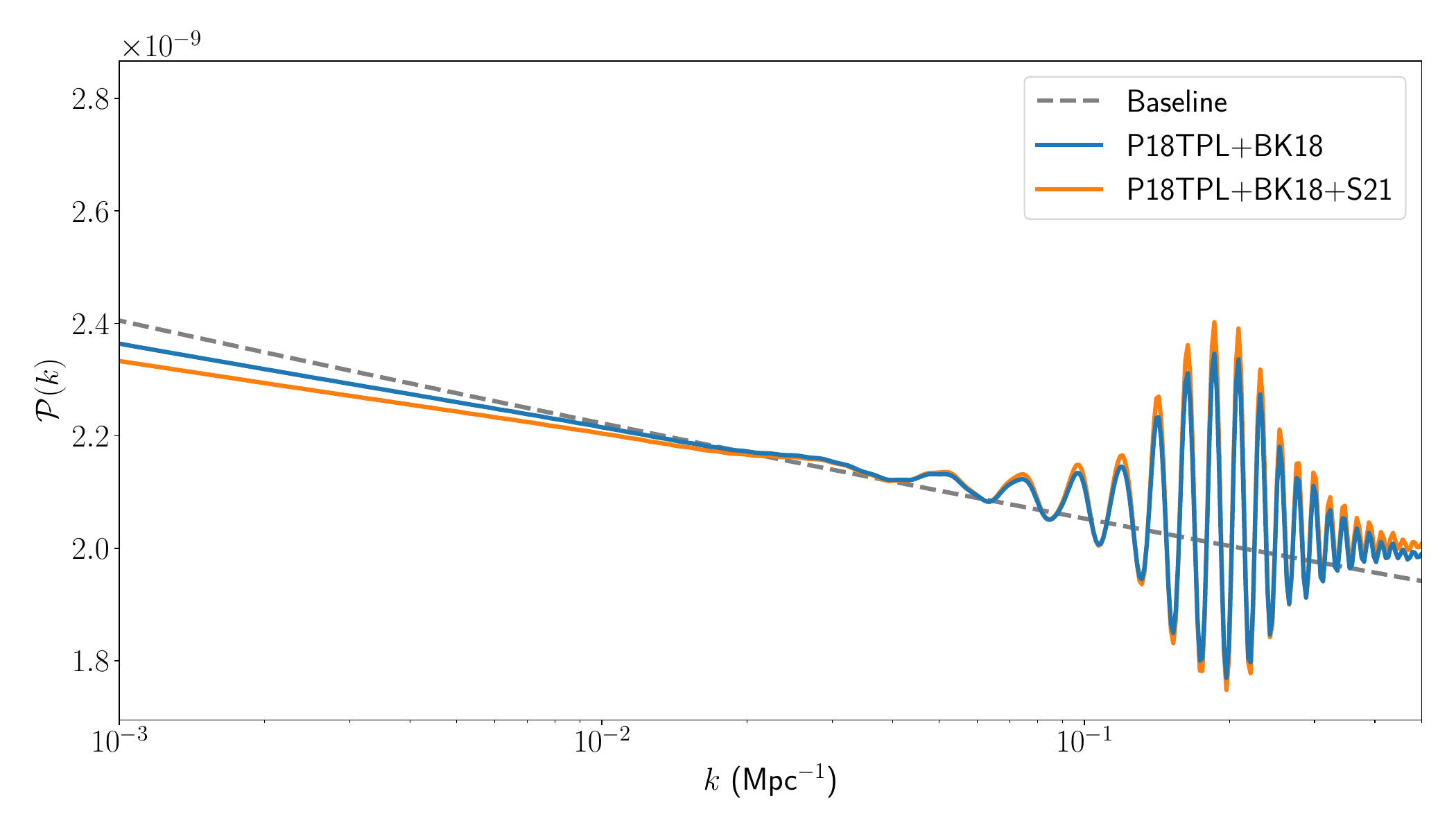}
\caption{\label{fig:PSK}Best fit power spectra obtained from different datasets combinations compared with the baseline spectra for the same dataset combinations. In all the cases, compared to the baseline, we find the spectral tilts move to bluer direction when we have fast roll dynamics. With $H_0$ prior from SH0ES results added, the redness of the tilt further decreases and the oscillations remain with located in the same scales with marginally higher magnitudes (within the Planck probed cosmological scales). Polarization and thereafter lensing addition to the temperature data restricts the shift in the tilt while keeping the oscillations largely unaltered. This particular best fit candidate, for P18TE provides 2.2 improvement in fit compared to power law. We find no improvement in fit to P18EE data. Similar improvements obtained in P18TEEE (2.7) data, however when SH0ES data is added we do not find any additional improvement in P18TEEE+BK18+S21. The reconstructed \textit{One Spectrum} from~\cite{Hazra:2022rdl} is plotted in the top left panel and the best fit spectrum for superimposed oscillations with a Gaussian envelope~\cite{Planck:2018inf} to P18TP is plotted in bottom left.}
\end{figure*}

\begin{figure*}[!htb]
\centering
\includegraphics[width=\textwidth]{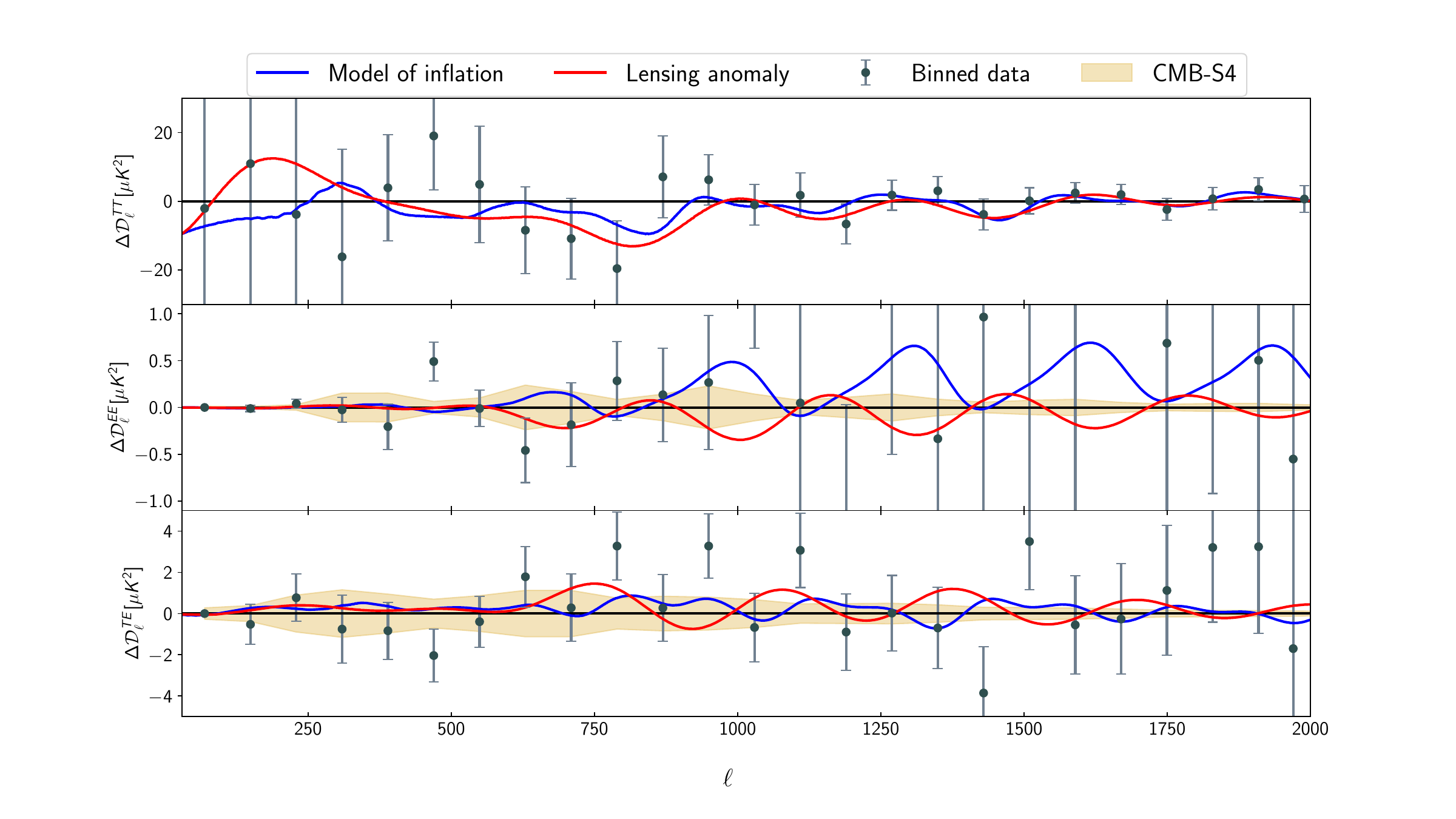}
\caption{\label{fig:Residual} The best fit angular power spectra, residual to the baseline best fit are plotted. Plik binned data, residual to the baseline best fit spectra are also plotted for comparison. In the residual panel for TT, we find that similar to the excess lensing, our model of fast roll fits the data better than the baseline (denoted by 0 line), at multipoles higher than 500. Interestingly we find that at higher multipoles the physics of fast roll and the lensing anomaly have completely different, out of phase signals in the E-mode auto-correlation spectrum. TE panel shows certain differences as well at different multipoles. The projected error band for EE and TE are plotted for CMB-S4 type observation.~\cite{S4}.}
\end{figure*}

In~\autoref{tab:chisq} we tabulate the main results of our analysis for the 8 data combinations. The best fit primordial power spectra for baseline and for the model with fast roll are plotted in~\autoref{fig:PSK}. These best fits are obtained with BOBYQA~\cite{BOBYQA} with the initial search region directed by the best fits obtained from the MCMC. Our model provides upto 11 improvement in fit to the data \textit{w.r.t.} the baseline. Most of the improvement (8.3) comes from the Planck 2018 temperature data. 
The features in the primordial power spectrum, introduced by the brief fast roll phase are located mainly between $k\sim0.1-0.2~{\rm Mpc}^{-1}$. Note that beyond $k\sim0.2~{\rm Mpc}^{-1}$, we have weak constraints from Planck. These oscillations are similar to superimposed oscillations with a Gaussian envelope \cite{Planck:2018inf} (plotted in the bottom left panel of \autoref{fig:PSK}) and to the 
exact reconstruction of the primordial spectrum \textit{One Spectrum} (plotted in top left panel of \autoref{fig:PSK}), that mimics $A_L$ discussed in~\cite{Hazra:2022rdl}.
The form of the oscillations is an ordered combination of sharp, resonant and sharp features that take the form of $\sin$, $\sin\log$ and $\sin$ respectively. The improvement in the fit mainly comes from the temperature power spectrum at high multipoles ($\ell\sim500-2000$) as demonstrated in~\autoref{fig:Residual}. Note that the primordial standard clock models~\cite{Chen:2011SC1,Chen:2014SC,Braglia:2021SC1,Braglia:2021SC2,Braglia:2021SC3} (sharp and resonant features) and Wiggly Whipped Inflation (sharp features)~\cite{Hazra:2021WWI} also provide similar improvement in fit while fitting the outliers at somewhat larger scales ($\ell<1200$ and $\ell<1000$ respectively).
We find that the $A_L$ effect in the temperature spectrum is mimicked by the effects of fast roll in the inflationary dynamics. Interestingly, they exhibit completely opposite effects in E-mode auto-correlation spectrum. Since at high multipoles, EE spectrum has a very low signal-to-noise ratio, it does not help in distinguishing these two types of features. Future ground based observations, such as Simons Observatory~\cite{SO}, CMB-S4~\cite{S4} complemented on large angular scales by LiteBIRD ~\cite{LiteBIRD:2022cnt} or cosmic variance limited CMB space proposals such as PICO~\cite{PICO} or CMBBHARAT~\cite{CMBBHARAT} will be able to provide the litmus test for these beyond standard model physics. We find marginally better results (10.7 improvement to the TP+BK18 data), with the physics of fast roll compared to the excess lensing effect (9.7 improvement~\cite{Planck:2018param}).

\begin{figure*}[!htb]
\centering
\includegraphics[width=0.495\columnwidth]{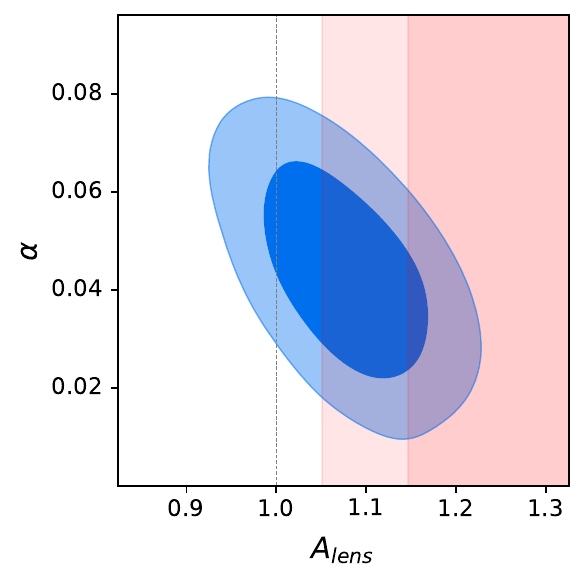}
\includegraphics[width=0.495\columnwidth]{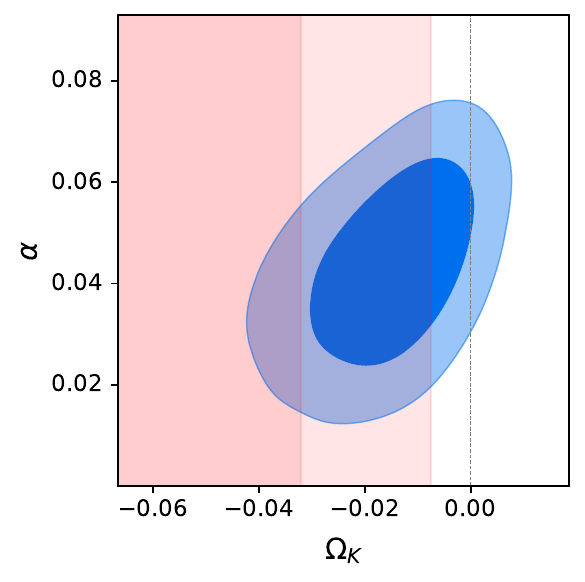}
\includegraphics[width=0.495\columnwidth]{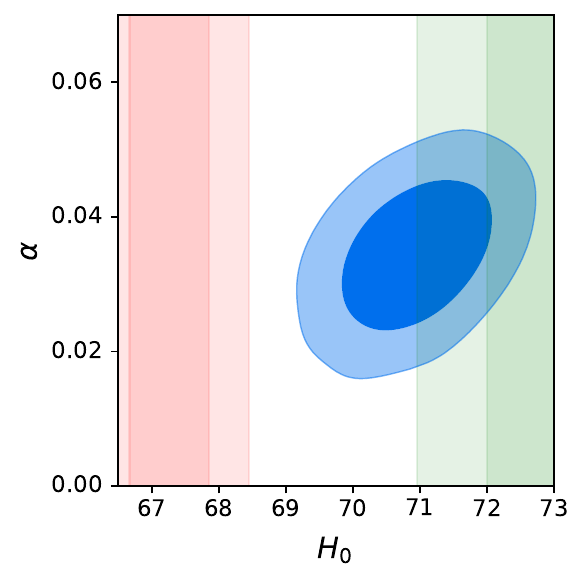}
\includegraphics[width=0.495\columnwidth]{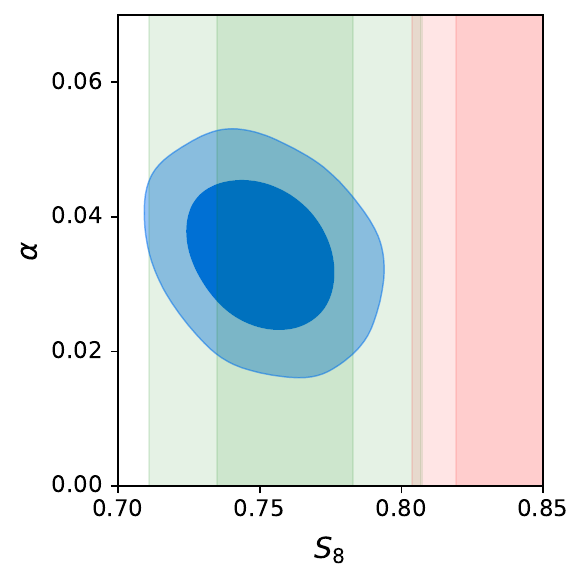}
\caption{\label{fig:degeneracy} Correlations between fast roll amplitude during inflation and other cosmological parameters in P18TT+BK18 data. The amplitude $\alpha$, is negatively correlated with $A_{\rm L}$ and positively correlated with the curvature density. The red bands indicate the P18TT bounds on these parameters in the baseline scenario. Note that the fast roll successfully brings $A_{\rm L}$ back to 1 and also make the flat Universe completely consistent with the data. The fast roll also helps in increasing the value of Hubble constant and decrease the value of $S_8$. The red bands indicate the baseline bounds while the green band in the $H_0$ plot indicate the constraints from SH0ES and in the $S_8$ plot shows the constraints from KIDS1000~\cite{KIDS1000}.}
\end{figure*}

\begin{figure*}[!htb]
\centering
\includegraphics[width=1.95\columnwidth]{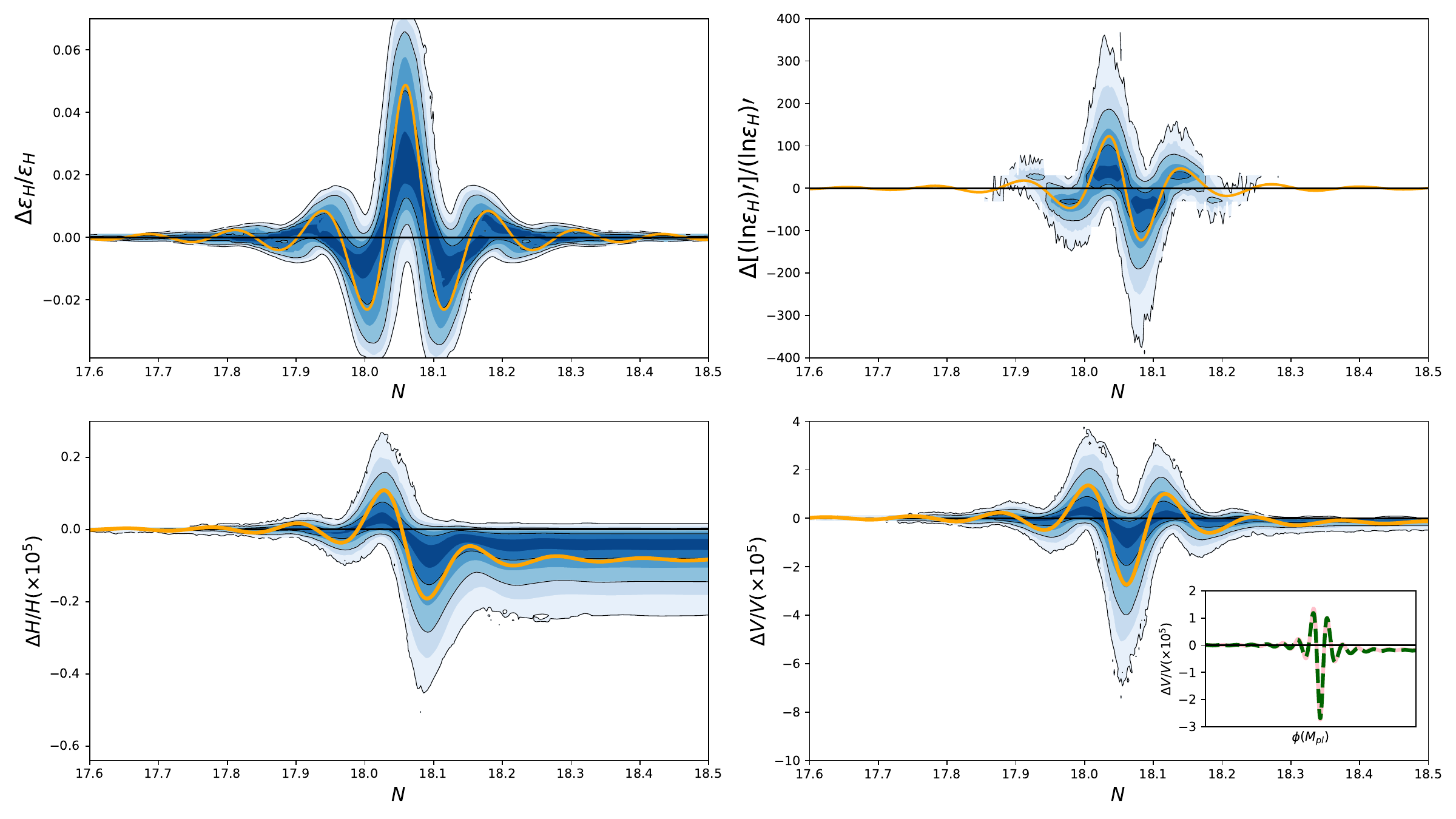}
\caption{\label{fig:evol-TP}Bounds on inflationary dynamics and reconstruction of the potential obtained from P18TP+BK18+S21 combination. In the top we plot the bounds on the fractional change in the Hubble flow functions ($\epsilon_H(N)$ and $d\ln\epsilon_H(N)/dN$) as a function of \textit{e-folds} \textit{w.r.t.} the baseline. The fractional change in the Hubble parameter during inflation is plotted to the bottom left. The change in the reconstructed potential is plotted in the bottom right. In all three plots the inward to outward bounding lines represent 1-3$\sigma$ confidence contours. The orange lines represent the best fit dynamics. The plots suggest a deviation from the slow-roll phase for about 0.5 \textit{e-fold}. In the inset of the bottom right plot, we provide the fractional change in the reconstructed best fit potential \textit{w.r.t.} a slow roll potential, as a function of scalar field. On top of that we plot an analytical fitting function for the potential.}
\end{figure*}

The correlations between the fast roll and the cosmological parameters are plotted in~\autoref{fig:degeneracy}. $A_{\rm L}$ extension of baseline model is correlated with the fast roll, as discussed for the localized linear oscillations in \cite{Planck:2018inf}. The degeneracy between $A_{\rm L}$ and the curvature and the tension with other datasets have been discussed in detail in~\cite{DiValentino:2019}. The threefold degeneracy was demonstrated in~\cite{Hazra:2022rdl} in the context of oscillations in the primordial power spectrum. First two figures in this plot show the degeneracy between lensing amplitude, curvature density and the fast roll dynamics. The fast roll amplitude $\alpha$ is negatively correlated with the lensing amplitude. We find that a fast rolling inflaton is able to bring the lensing amplitude (represented by $A_{\rm L}=1$) back within the 1$\sigma$ C.L. Similarly we find that the fast roll is positively correlated with the curvature density of the Universe that brings back the flat Universe scenario within the 1$\sigma$ contour. 
In these two plots we have used P18TT+BK18 data and we have also fixed the nuisance parameters, optical depth and $\omega$ to their best fit values. 

The bounds on the parameters quoted in~\autoref{tab:chisq} highlight the degeneracy between a few other cosmological parameters. Importantly, we find that with the improved fit to the data, the intermediate fast roll also helps in shifting the posterior distribution of $H_0$ to higher values, as happens for $A_L$. For temperature data, we find a shift in the mean value of $H_0$ by 1.3$\sigma$ with only marginal increase in the uncertainty compared to the baseline scenario. While earlier analyses~\cite{Planck:2018inf,Liu:2019dxr} reported that certain type of features in the primordial spectrum decrease the inference of $H_0$ from CMB (that increases the tension with the local Hubble measurements), our model, interestingly ameliorates the tension. This decrease in the tension allows us to include $H_0$ prior from the recent SH0ES~\cite{Riess:2021jrx} in the analysis. For the combined datasets P18TT+BK18+S21 and P18TP+BK18+S21, an intermediate fast roll in the dynamics improves the fit to the data by a $\Delta\chi^2$ of 19-20 compared to the baseline. About 11 improvement in fit comes from the CMB data and nearly equal improvement is obtained from the $\chi^2$ from SH0ES $H_0$ prior. The $\chi^2$ improves as the baseline does not allow $H_0$ to increase without worsening the fit to the CMB data.
We would like to highlight that the $\chi^2$ from CMB in the joint P18TT+BK18+S21 and P18TP+BK18+S21 analyses is better than the baseline fit to the CMB-\textit{only} datasets with $H_0\sim70-71$. We also find that the feature parameters in the bestfits for P18TT+BK18 and for P18TEEE+BK18 are different, 
but become very close once S21 is fold in. The correlation between fast roll and the Hubble parameter can be noticed in the third plot from the left in~\autoref{fig:degeneracy}. The contour is plotted from the analysis with P18TT+BK18+S21 data. We find a significant shift in $H_0$ towards its locally measured value with a fast rolling inflaton.

The scalar tilt is strongly correlated with the $H_0$ in the CMB temperature data. In the table the increase in $n_s$ (approximately $1-2\epsilon_1-\epsilon_2$) can be noticed compared to the baseline values. Here too we only find a marginal increase in the uncertainty compared to the baseline. The extent of increase reduces when the polarization data are added but at the same time it decreases the uncertainty in the estimation. This keeps the increase in parameter similarly significant as we find in the temperature data. When lensing data is added, we find that the improvement in fit reduces in both P18TPL+BK18 and P18TPL+BK18+S21. Since lensing data, combined with temperature and polarization data restricts the spectral tilt to a lower value, it further reduces the mean value of $H_0$. The change in the scalar tilt in the best fit fast roll model compared to the baseline best fit can be noticed in the different data analysis plots in~\autoref{fig:PSK}. 

The intermediate fast roll, apart from providing \textit{a physical solution} to the lensing anomaly problem in a flat Universe with a higher Hubble constant, also helps in reducing the matter density and the $\sigma_8$ normalization parameter (with direct reconstruction a numerical solution was discussed in\cite{HazraHST:2018,Keeley:2020}). A combination of these two parameters, $S_8=\sigma_8\sqrt{\Omega_m/0.3}$, is found to be lower than the baseline value when the inflationary dynamics has an intermediate fast roll phase. Without any prior on the Hubble constant, we find (as quoted in~\autoref{tab:chisq}) 1$\sigma$ decrease in $S_8$ mean compared to the baseline value in the analysis with temperature data. This decrease makes the CMB data more compatible with the galaxy clustering and weak lensing measurements from DES~\cite{DES} and KiDS~\cite{KIDS1000} and brings in a new concordance. Use of the prior on Hubble parameter further improves this agreement. Therefore the parameter $\alpha$ for the fast roll is anti-correlated with $S_8$. This is a very interesting feature compared to other representatives of new physics discussed in the context of Hubble tension, such as  early dark energy~\cite{PoulinPRL} or some models of scalar-tensor and early modified gravity theories ~\cite{Rossi:2019lgt,Ballardini:2020iws,Braglia:2020auw} where $S_8$ increases with $H_0$. 

\autoref{tab:chisq} provides the confidence limits for the detection of the intermediate fast roll. The number in the C.L. column of the table mainly indicates the confidence limits of the parameter $\alpha$ deviating from its baseline value $\alpha=0$. In all analyses without the prior on Hubble parameter, we find between 1-2$\sigma$ preference for the fast roll. A prior on Hubble parameter increases the significance to beyond 3$\sigma$ for P18TT+BK18+S21 data. Polarization data marginally decreases the significance. When lensing data is added, it decreases the significance further to 2$\sigma$.  

From the analysis chains, we estimate the posterior of the Hubble flow functions and the derived potential for the scalar field. We use {\tt fgivenx}~\cite{fgivenx} package to plot the confidence bands on the functions.The bounds on the inflationary dynamics are plotted in~\autoref{fig:evol-TP} for P18TP+BK18+S21. We highlight fractional change in the dynamics during the intermediate fast roll phase only. The significant part of the fast roll phase lasts for 0.5 \textit{e-folds}. For P18TP+BK18+S21, we notice the \textit{zero} line for the baseline is falling within 2-3$\sigma$. The change in the Hubble parameter integrated from $\epsilon_H(N)$ can be distinguished in 2 parts, decaying oscillations and a step function. Therefore, when the potential is reconstructed using $V(\phi[N])=3M_{\rm Pl}^2H[N]^2(1-\epsilon_H[N]/3)$, we find these two features in the inflationary potential. Being agnostic about the baseline potential, we plot the numerically evaluated potential from the Hubble flow functions and a closely matching analytical approximation
\begin{equation}
    \frac{\Delta V(\phi)}{V_{\rm baseline}(\phi)} = \frac{\alpha\cos[\omega(\phi-\phi_0)]}{1+\beta(\phi-\phi_0)^2} \,, 
\end{equation}
in the same plot. 


\paragraph{\textit{Conclusions}:}
In this \textit{Letter} we have provided templates in the Hubble flow function during inflation leading to primordial features which mimic the effect of $A_{\rm L}$ in the Planck 18 data. This construction gives theoretical support to localized linear oscillations proposed in \cite{Planck:2018inf}
and \textit{One Spectrum}~\cite{Hazra:2022rdl} obtained through the reconstruction of the primordial spectrum.
We summarize our findings below.
\begin{enumerate}
    \item \textit{An intermediate fast roll in the inflationary dynamics for a period of 0.5 \textit{e-folds} is able to solve the lensing anomaly in a flat Universe}. This fast roll has a completely different signature in the polarization anisotropy spectrum compared to $A_L$, which can be distinguished by more precise CMB polarization measurements at high multipoles.

    \item This intermediate fast roll provides nearly 11 improvement in fit compared to the standard baseline model when compared with Planck and BICEP/Keck 2018 data. Most of the improvement (more than 8) comes from temperature data. The model is supported at between 1-2$\sigma$ C.L. when \textit{only} CMB data is used.
    
    \item Importantly, this fast roll model \textit{simultaneously} prefers a higher value of $H_0$ and lower value of $S_8$ (only with CMB) and matter density which brings the CMB closer to the local Hubble measurements and galaxy weak lensing measurements. When priors on Hubble parameter is used with CMB, we find nearly or more than $3\sigma$ significance in different combinations of datasets. With better agreement to both CMB and SH0ES data, our model provides nearly 20 improvement in fit compared to baseline.   
    
    \item A reconstruction of inflaton potential from the Hubble flow function indicates a damped oscillatory modification to the baseline slow roll potential.   
\end{enumerate}
    
The intermediate fast roll phase in the inflation dynamics, apart from its unique signature in the CMB anisotropy power spectra, also imprints signals in the large scale structure, as studied for other types of primordial features \cite{Hazra:2012LSS,Ballardini:2016hpi,Ballardini:2019tuc}. The characteristic frequency of the superimposed oscillations could be detected in galaxy correlation function or in the power spectrum alongside the baryon acoustic oscillation bump with the ongoing and upcoming LSS probes such as DESI~\cite{DESI}, LSST~\cite{LSST}, Euclid~\cite{Euclid}. 

\begin{acknowledgments}
The authors acknowledge the use of computational resources
at the Institute of Mathematical Science’s High Performance Computing facility [Nandadevi]. DKH would like to thank Matteo Braglia and Xingang Chen for important discussions. AA, FF, DKH acknowledge travel support through the India-Italy mobility program. FF acknowledges financial support from the contract by the agreement n. 2020-9-HH.0 ASI-UniRM2 
``Partecipazione italiana alla fase A della missione LiteBIRD". AS would like to acknowledge the support by National Research Foundation of Korea NRF-2021M3F7A1082053, and the support of the Korea Institute for Advanced Study (KIAS) grant funded by the government of Korea. 

\end{acknowledgments}

\bibliography{main}

\end{document}